\documentclass[aps,prd,twocolumn,nofootinbib,amsmath,amssymb,showpacs]{revtex4-1}

\usepackage{graphics}
\usepackage[usenames,dvipsnames]{color}
\usepackage{hyperref}
\usepackage{mwe}
\usepackage[subnum]{cases}
\usepackage{lipsum}

\begin{document}
 
\title{Radial oscillations of quark stars from perturbative QCD}

\author{Jos\'e C. {\sc Jim\'enez}}
\author{Eduardo S. {\sc Fraga}}

\affiliation{Instituto de F\'\i sica, Universidade Federal do Rio de Janeiro,
Caixa Postal 68528, 21941-972, Rio de Janeiro, RJ, Brazil}

\date{\today}


\begin{abstract}

We perform the general relativistic stability analysis against adiabatic radial oscillations of unpaired quark stars obtained using the equation of state for cold quark matter from perturbative QCD, the only free parameter being the renormalization scale. This approach consistently incorporates the effects of interactions and includes a built-in estimate of the inherent systematic uncertainties in the evaluation of the equation of state. We also take into account the constraints imposed by the recent gravitational wave event GW170817 to the compact star masses and radii, and restrict their vibrational spectrum.
  
\end{abstract}



\maketitle


\section{Introduction}

Pulsar observations have shown that neutron stars (NS) must suffer different kinds of mechanical deformations, e.g. radial and non-radial oscillations, rotation and glitches, along their lifetime in order to reproduce radio, X-ray, gamma-ray and other electromagnetic signatures \cite{Shapiro:1983,Glendenning:2000,Haensel:2007,Benhar:2004xg,Doneva:2015jba,Haskell:2015jra,Harding:2006qn,Watts:2006hk}. Recently, the LIGO and Virgo observatories measured gravitational waves coming from the merger of neutron stars, the GW170817 event, opening a new window to probe NS responses to some of the disturbances produced by tidal deformations in the inspiral phase \cite{TheLIGOScientific:2017qsa}. Besides the usual constraints -- e.g.  the existence of $\sim 2 M_\odot $ neutron stars \cite{Demorest:2010bx,Antoniadis:2013pzd,Kurkela:2014vha,Buballa:2014jta} and tidal deformabillities \cite{Annala:2017llu,Rezzolla:2017aly,Most:2018hfd,Abbott:2018exr} --, mechanical responses could potentially provide signatures of the presence of quark matter in their cores \cite{Most:2018eaw,Wei:2018tts,Tanimoto:2019tsl,Annala:2019puf} or indicate the presence of strange quark stars \cite{Bauswein:2008gx,Bauswein:2009im}.

In this work we compute the radial pulsation frequencies and periods of unpaired quark stars coming from possible radial perturbations occurring at different stages of the pulsar's lifetime\footnote{Compact star oscillations can happen due to several reasons, e.g. accretion from a partner in a binary system or due to impact of interstellar objects as asteroids or comets \cite{Geng:2015vza,Dai:2016qem}.}. The general relativistic radial pulsation analysis framework was designed long ago by Chandrasekhar \cite{Chandrasekhar:1964zza}, being applied initially only to polytropic equations of state (EoSs), and only much later to more realistic nuclear EoSs \cite{Glass:1983}. Many radial oscillation modes were calculated for modern sets of EoSs for cold NS by Kokkotas and Ruoff \cite{Kokkotas:2000up}. 
	
For quark stars, radial pulsations were analyzed mostly using the MIT bag model to build the equation of state for cold quark matter \cite{Benvenuto:1989kc,Benvenuto:1991,Vath:1992,Lugones:1995vg,Benvenuto:1998tx,VasquezFlores:2010eq}, in some cases adding constant corrections to the strange quark mass and interactions which behave effectively only as being of long range. Results suggested that the periods of the fundamental mode were very low to be detected \cite{Benvenuto:1991jz,Horvath:1991pv}, which motivated the search for the so-called non-radial oscillations that would have higher periods which could be measured through gravitational wave observations \cite{Flores:2013yqa,Flores:2018pnn}. 

In the following, we adopt an equation of state for cold quark matter extracted from perturbative quantum chromodynamics (pQCD) and impose recent gravitational wave constraints for compact star masses and radii \cite{Rezzolla:2017aly,Most:2018hfd} to compute the fundamental and first excited mode frequencies ($n=0,1$) of unpaired quark stars. This approach consistently incorporates the effects of interactions and includes a built-in estimate of the inherent systematic uncertainties in the evaluation of the equation of state, so that in several cases we present bands instead of lines. This goes beyond the misleading precision of the MIT bag model description, providing a more realistic range of possibilities, in a framework that can be systematically improved.

To study the radial pulsations of quark stars, we use the framework of Ref. \cite{Gondek:1997fd}, where the original Sturm-Liouville form is turned into a pair of first-order coupled differential equations. For the equation of state, we use the pressure for cold QCD matter of Ref. \cite{Kurkela:2009gj}, which can be cast in a pocket formula as described in Ref. \cite{Fraga:2013qra}\footnote{Cold perturbative QCD has a long story \cite{Freedman:1976ub,Freedman:1977gz,Baluni:1977ms,Toimela:1984xy,
Blaizot:2000fc,Fraga:2001id,Fraga:2001xc,Fraga:2004gz,Kurkela:2009gj,Fraga:2013qra} and, although its realm of validity corresponds to much higher densities, it is relevant in modelling the equation of state of compact stars, since QCD short-range interactions become important at intermediate densities, reachable in the interior of NS \cite{Fraga:2001id,Fraga:2001xc,Fraga:2004gz,Kurkela:2009gj,Fraga:2013qra}.}. 

This work is organized as follows. In Sec. \ref{sec:setup} we set up the formalism needed to study radial pulsations of quark stars, i.e., the cold quark matter equation of state and the relativistic oscillation equations. In Section \ref{sec:QManalysis} we present our results for the stability of quark stars. Section \ref{sec:conclusion} presents our summary and final remarks.

\section{SETUP}
	\label{sec:setup}
\subsection{EoS for cold unpaired quark matter}
	\label{Subsec:QCDmatter}

Since we are interested in the physics of quark stars, we adopt an EoS for cold quark matter obtained from cold and dense perturbative QCD, satisfying $\beta$-equilibrium and electric charge neutrality. This equation of state was computed up to second order in the strong coupling $\alpha_{s}$ by Kurkela et al. \cite{Kurkela:2009gj}, including the effects of the renormalization group on $\alpha_{s}(\bar{\Lambda})$ and the strange quark mass $m_{s}(\bar{\Lambda})$. As usual, the perturbative calculation brings about an additional scale, the renormalization scale $\bar{\Lambda}$, a parameter that has to be varied within some range, and can be constrained by the phenomenology \cite{Fraga:2001id}. This provides an estimate of the inherent systematic uncertainties in the evaluation of the equation of state.

The full result of Ref. \cite{Kurkela:2009gj} can be cast into the following pocket formula, \cite{Fraga:2013qra} which we call FKV:
	\begin{equation}
	P_{\rm QCD}=P_{\rm SB}(\mu_{B})\left(c_{1}-\frac{a(X_{\rm FKV})}{(\mu_{B}/{\rm GeV})-b(X_{\rm FKV})}\right)  \; ,
	\label{pressureFKV}
	\end{equation}	
where $P_{\rm SB}$ represents the Stefan-Boltzmann gas. This formula includes the contributions from massless up and down quarks, a strange quark with running mass, and massless electrons. It is in $\beta$-equilibrium and electrically neutral. Here, $\mu_{B}$ is the baryon chemical potential and we use the dimensionless version of the renormalization scale, $X_{\rm FKV}=3\bar{\Lambda}/\mu_{B}$, which can vary between $1$ and $4$, as discussed in Ref. \cite{Kurkela:2009gj}. The auxiliary functions that enter this pressure are defined as
	\begin{equation}
	a(X_{\rm FKV})=d_{1}X^{-\nu_{1}}_{\rm FKV},\hspace{0.5cm}b(X_{\rm FKV})=d_{2}X^{-\nu_{2}}_{\rm FKV},
	\end{equation}		
	with the following fit values (for details, see Ref. \cite{Fraga:2013qra})
	\begin{equation}
	c_{1}=0.9008,\hspace{0.2cm}d_{1}=0.5034,\hspace{0.2cm}d_{2}=1.452,
	\end{equation}
	\begin{equation}
	\nu_{1}=0.3553, \hspace{0.3cm}\nu_{2}=0.9101.
	\end{equation}

\begin{figure*}[t]
\begin{center}
\hbox{\includegraphics[width=0.5\textwidth]{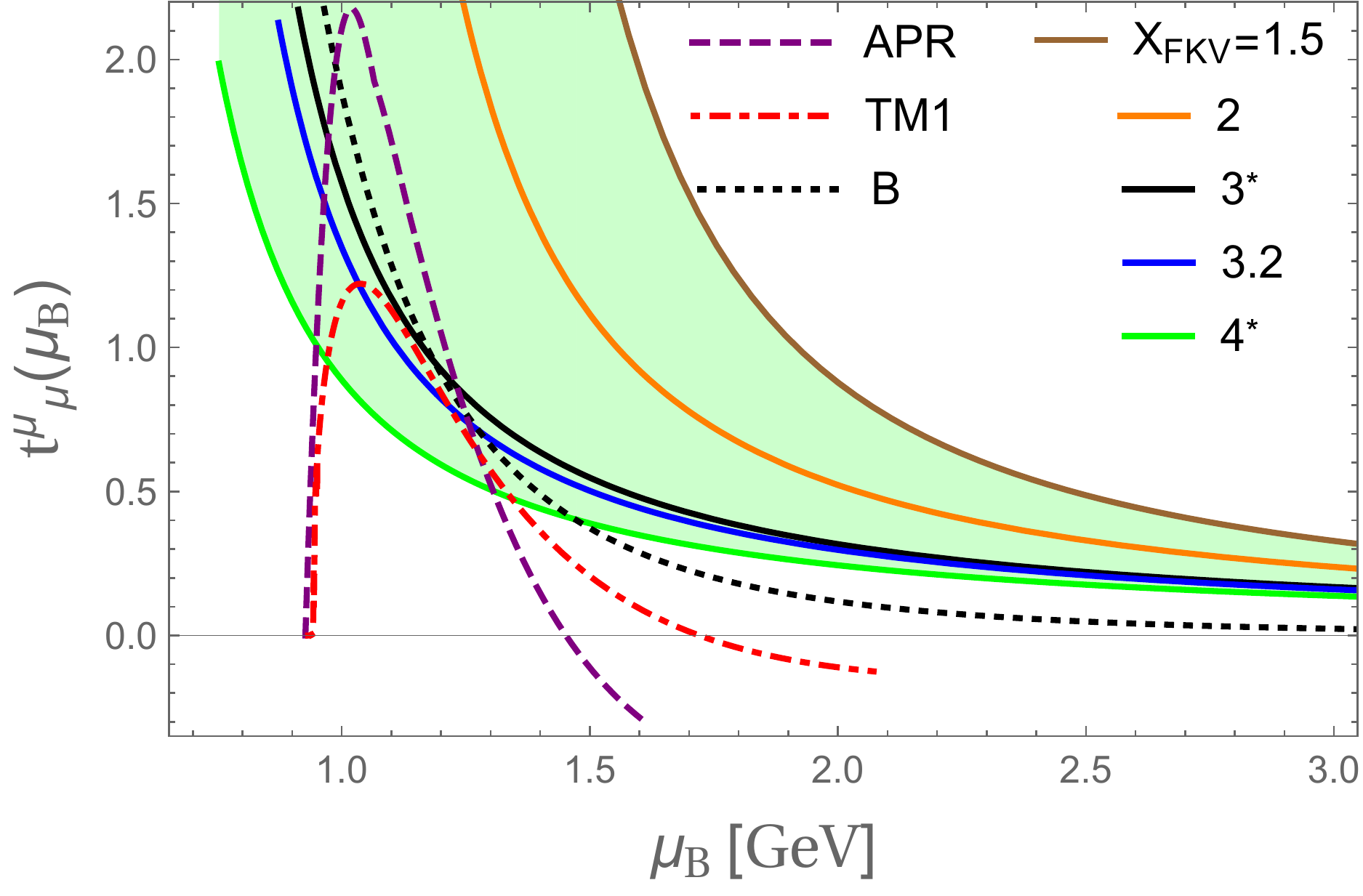}
	  \includegraphics[width=0.5\textwidth]{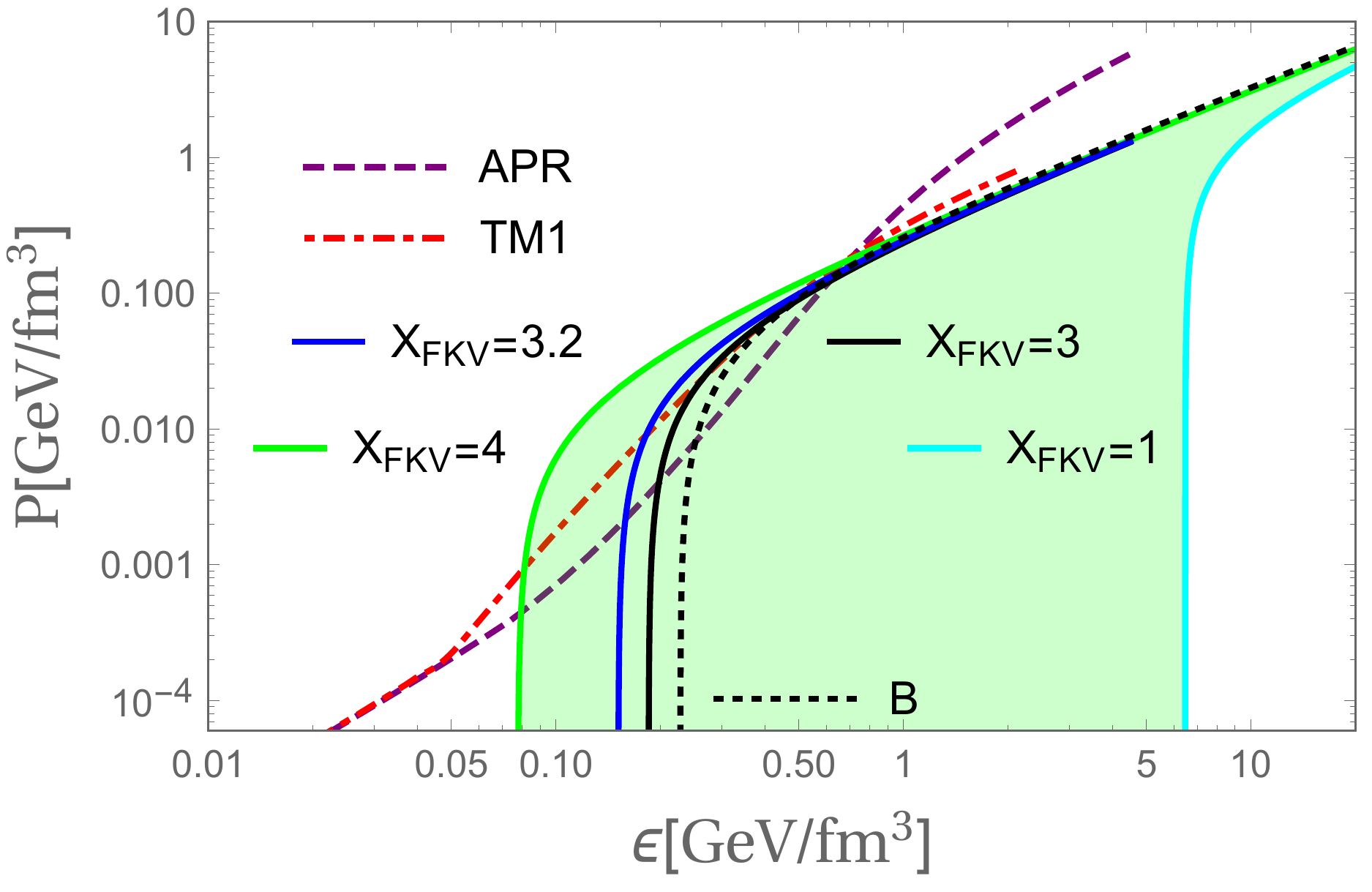}}
\caption{\textit{Left panel:} Trace anomaly for the cold pQCD result of Refs. \cite{Kurkela:2009gj,Fraga:2013qra} normalized by the Stefan-Boltzmann pressure as a function of the baryon chemical potential for different values of $X_{\rm FKV}$ (continuous) and for the bag model with $B=(145\rm MeV)^4$ (dotted). We mark with asterisks the cases of $X_{\rm FKV}$ between 3 and 4 since only within this range one obtains at least two-solar mass stars. For comparison, we also show the behavior of the trace anomaly for two well-known nuclear matter equations of state, APR and TM1 (see text). \textit{Right panel:} Equations of state, $P=P(\epsilon)$, for a few values of $X_{\rm FKV}$ and the bag model $B$, to be discussed in the next section. We also show the nuclear matter APR and TM1 equations of state.}
\label{fig:AnomAndEoSs}
\end{center}
\end{figure*}

From Eq. (\ref{pressureFKV}) one can easily compute the perturbative trace anomaly of QCD  normalized by the Stefan-Boltzmann gas, obtaining
   \begin{equation}
   t^{\mu}_{\mu}(\mu_{B}, X_{\rm FKV})=\frac{\mu_{B}}{\rm GeV}\frac{a(X_{\rm FKV})}{\left[(\mu_{B}/{\rm GeV})-b(X_{\rm FKV})\right]^{2}} \; ,
   \end{equation}
which gives a measure of the role of interactions encoded in the breaking of conformal symmetry \cite{Andersen:2011ug}. In Fig. \ref{fig:AnomAndEoSs} we show the pQCD normalized trace anomaly for different values of $X_{\rm FKV}$, and compare it to the result obtained from the bag model for $B=(145\rm MeV)^4$, which vanishes very quickly with $\mu_{B}$. For comparison we also plot in this figure the trace anomalies of standard nuclear matter EoSs, i.e. the ones from Akmal et al. \cite{Akmal:1998cf} (dubbed APR\footnote{This EoS is also known as APR4 or A$18+\delta{v}+{\rm UIX}^{*}$.}) and Shen et al. \cite{Shen:1998gq} (dubbed TM1), to be used later. For the perturbative case, we show a band that represents a measure of the actual uncertainties, to be contrasted to the apparent, misleading precision of the bag model line.

Strange stars are quark stars that are self bound by QCD interactions and satisfy the Bodmer-Witten hypothesis of strange quark matter being the true ground state of nuclear matter in the vacuum. Strange matter configurations at zero temperature and zero pressure would have $E/A=\epsilon_{\rm QCD}/n_{B}<930$MeV, i.e. energy per baryon lower than iron-56 \cite{Glendenning:2000}.   
   
The FKV formula has a broad parameter space which allows for the existence of configurations of self-bound matter. It was shown in Ref. \cite{Kurkela:2009gj} that this condition is satisfied for values of $X_{\rm FKV} \sim 3-4$. A similar conclusion was obtained within the Nambu--Jona-Lasinio (NJL) model for quark matter for the most common parametrizations of the EoS \cite{Buballa:2003qv}. In what follows, we consider only bare quark stars, i.e. stars without a nuclear crust, which depending on the value of $X_{\rm FKV}$ will be self-bound stars or ordinary quark stars.
	
A technical detail one has to keep in mind when using the FKV pocket formula for the pressure and in obtaining other thermodynamic quantities is that, since $P_{\rm QCD}$ is a function of $\mu_{B}$ and $X_{\rm FKV}$, when obtaining the energy density $\epsilon_{\rm QCD}$ one has the freedom to choose first a fixed value of $X_{\rm FKV}=X_{\rm FKV}(\mu_{B})$ (e.g., $X_{\rm FKV}=2$, i.e., $\bar{\Lambda}=(2/3)\mu_{B}$) in $P_{\rm QCD}$ and then build the energy density using the thermodynamic relation $\epsilon_{\rm QCD}=-P_{\rm QCD}+n_{B}\mu_{B}$, or to consider $X_{\rm FKV}$ an independent constant and then build $\epsilon_{\rm QCD}$ by taking derivatives in $\mu_{B}$ keeping $X_{\rm FKV}$ constant when computing the baryon number density $n_{B}$ (see, e.g., Ref. \cite{Haque:2014rua} for a similar discussion at high temperatures). In this work, we have chosen to first fix the values of $X_{\rm FKV}$ as functions of $\mu_{B}$, and then build the EoS, $P_{\rm QCD}=P_{\rm QCD} (\epsilon_{\rm QCD})$. Our approach differs from the other option by a few percent for low values of $\mu_{B}$. 

\subsection{Linear radial oscillations of quark stars}
  \label{Subsec:radial}
  
Below we summarize the main aspects concerning the stability analysis of linear radial oscillations (pulsations) in quark stars\footnote{Non-linear radial oscillations could be important only for stellar configurations around the maximum mass, producing unstable modes. For more details, see Ref. \cite{Gourgoulhon:1995jjj}.}. For simplicity, we assume that the stars are static and spherically symmetric, so that one can use the Schwarzschild-like line element, having as non-trivial metric functions $e^{\nu(r)}$ and $e^{\lambda(r)}$, for the temporal and radial parts, respectively. Then, Einstein's equations are solved for stellar configurations in hydrostatic equilibrium, yielding the Tolman-Oppenheimer-Volkov (TOV) equations (using $G=c=1$) \cite{Glendenning:2000,Shapiro:1983}
  \begin{equation}
  \frac{dP}{dr}=-\frac{\epsilon\mathcal{M}}{r^{2}}\left(1+\frac{P}{\epsilon}\right)\left(1+\frac{4\pi{r^3}{P}}{\mathcal{M}}\right)
  \left(1-\frac{2\mathcal{M}}{r}\right)^{-1} \; ,
   \label{TOV1}
  \end{equation}
  \begin{equation}
  \frac{d\mathcal{M}}{dr}=4\pi{r}^{2}\epsilon \; ,
  \label{TOV2}
  \end{equation}
  \begin{equation}
  \frac{d\nu}{dr}=-\frac{2}{P+\epsilon}\frac{dP}{dr} \; ,
  \label{TOV3}
  \end{equation}
where $P$ is the pressure, $\epsilon$ is the energy density, and $\mathcal{M}$ is the gravitational mass inside the radius $r$. 

To solve Eqs. (\ref{TOV1})-(\ref{TOV2}), one needs the EoS, $P=P(\epsilon)$, as an input. Then, one imposes that at the origin $\mathcal{M}(r=0)=0$ and $P(r=0)=P_{0}$, and the integration must end when $P(r=R)=0$, i.e. at the surface of the star, its total mass being $\mathcal{M}(r=R)=M$. Additionally, in order to solve Eq. (\ref{TOV3}) for $\nu$ we use the boundary condition $\nu(r=R)=\ln\left(1-{2M}/{R}\right)$. This ensures that this metric function $\nu(r)$ will match continuously the Schwarzschild metric outside the star, in agreement with Birkhoff's theorem \cite{Shapiro:1983}.
 
In order to obtain the equations for radial pulsations of relativistic stars, one begins by perturbing the spacetime and fluid components while preserving the spherical symmetry of the unperturbed star. These perturbations are introduced into Einstein's equations together with the energy-momentum and baryon number conservation laws, neglecting nonlinear terms. Historically, it was Chandrasekhar \cite{Chandrasekhar:1964zza} who first presented the second-order pulsating differential equations in the form of a Sturm-Liouville problem, which after being solved yields eigenvalues and eigenfunctions for the radial perturbations. Only decades later it was realized by Vath and Chanmugan \cite{Vath:1992} that these equations could be transformed into a set of two first-order differential equations by choosing appropriate variables, namely $\Delta{r}/r$ and $\Delta{P}/P$. Later, Gondek et al. \cite{Gondek:1997fd} found convenient to rewrite these equations for the relative radial displacement $\Delta{r}/r$ and the Lagrangian perturbation of the pressure $\Delta{P}$. This last set of equations is well adjusted to numerical techniques since one directly imposes the boundary condition at the star's surface. Moreover, these equations do not involve any derivatives of the adiabatic index, $\Gamma$, which is sensitive to the EoS being used.
 
Defining $\Delta{r}/r\equiv\xi$ and $\Delta{P}$ as the independent variables (omitting their harmonic time dependence, ${e}^{i\omega{t}}$) for the pulsation problem, one obtains the following system of equations (also with $G=c=1$)\cite{Gondek:1997fd}:
	\begin{equation}
	\frac{d\xi}{dr}=-\frac{1}{r}\left(3\xi+\frac{\Delta{P}}{\Gamma{P}}\right)-\frac{dP}{dr}\frac{\xi}{(P+\epsilon)} \; ,
	\label{Rad1}
	\end{equation}
	\begin{multline}
	\frac{d\Delta{P}}{dr}=\xi\left\lbrace{\omega^{2}e^{\lambda-\nu}(P+\epsilon)r-4\frac{dP}{dr}}\right\rbrace+ \\
	\xi\left\lbrace\left(\frac{dP}{dr}\right)^{2}\frac{r}{(P+\epsilon)}-8\pi{e^{\lambda}}(P+\epsilon)Pr\right\rbrace + \\
	\Delta{P}\left\lbrace{\frac{dP}{dr}\frac{1}{P+\epsilon}-4\pi(P+\epsilon)r{e}^{\lambda}}\right\rbrace \; ,
	\label{Rad2}
	\end{multline}	
where $\omega$ is the oscillation frequency.
	
The boundary conditions are given by the following:
	\begin{itemize}	
	 \item Physical smoothness at the center of the star requires that, when $r \to 0$, the coefficient of the $1/r$ term in Eq. (\ref{Rad1}) must vanish. Thus, we impose that
	\begin{equation}
	(\Delta{P})_{\rm center}=-3(\xi\Gamma{P})_{\rm center}.
	\label{BC1}
	\end{equation}

	\item Normalizing the eigenfunctions to $\xi(0)=1$ and knowing that $P(r \to R) \to 0$, we see that the Lagrangian perturbation in the pressure at the surface vanishes. Thus
	\begin{equation}
	(\Delta{P})_{\rm surface}=0.
	\label{BC2}
	\end{equation}
	\end{itemize}
	
	In order to solve simultaneously Eqs. (\ref{Rad1})-(\ref{BC2}) numerically, we use the following recipe: 
	\begin{itemize}	
	\item Solve the TOV equations for the EoS to be used in the analysis to calculate the coefficients of Eqs. (\ref{Rad1})-(\ref{Rad2}), i.e. combinations of $\Gamma(r)$, $P(r)$, $\epsilon(r)$, $\lambda(r)$, and $\nu(r)$ for a given central pressure.
	
	 \item After solving the desired equations with their boundary conditions and a set of trial values for $\omega^{2}$, obtain an oscillating behavior of $\Delta{P}$ and $\xi$ as functions of $\omega$.
	 
	 \item Only the discrete values of the frequency that satisfy $\Delta{P}(\omega^{2}_{i})=0$ are considered eigenfrequencies of the system\footnote{Our code reproduces the pulsation frequencies of Kokkotas and Ruoff \cite{Kokkotas:2000up}.}.
	\end{itemize}	
	
Although this procedure is different from the more commonly used Sturm-Liouville eigenvalue problem, it also examines the squared frequencies $\omega^{2}_{n}$ satisfying $\omega^{2}_{0}<\omega^{2}_{1}<\omega^{2}_{2}<\cdot\cdot\cdot$. If $\omega^{2}_{n}>0$, the frequency is real and the mode is stable and oscillatory. On the other hand, if $\omega^{2}_{n}<0$, then the frequency is purely imaginary and the mode is unstable. For the global stability of the star, it is sufficient to look only at the fundamental (lowest) eigenvalue, $\omega^{2}_{0}$. If $\omega^{2}_{0}>0$, then all $\omega^{2}_{n}>0$ and the star is stable. If $\omega^{2}_{0}<0$, then there is at least one unstable mode and the star becomes unstable. The transition between these two stellar states occurs when this fundamental frequency vanishes, i.e. $\omega_{0}=2\pi{f}_{0}\rightarrow{0}$ \cite{Shapiro:1983,Bardeen:1966aa}. Thus, when the fundamental mode reaches zero in $\Delta{P}(\omega^{2}_{0}=0)=0$, we obtain the maximal stable mass configuration for a given equation of state before gravitational collapse.	
		 
 We now calculate the last coefficient needed to solve the above pulsation equations, i.e. the adiabatic index $\Gamma$ of the quark matter system. Following Ref. \cite{Gondek:1997fd}, for cold matter the chemical reactions between components (quarks and electrons, in this case) are so slow ($\tau_{\rm reac}\gg\tau_{\rm dyn}$) that after perturbations the composition is not modified, which allows us to write for cold, adiabatic and isentropic (zero entropy) quark stars, the adiabatic index as
\begin{equation}
\Gamma{~\equiv~}\left[\left(1+\frac{\epsilon}{P}\right)\frac{\partial{P}}{\partial{\epsilon}}\right]_{s=T=0}   \; . 
\end{equation}
In the case of the bag model, it is easy to find the following analytic expression for this index:
\begin{equation}
\Gamma_{\rm MIT}=\frac{4}{3} \left( 1+ \frac{B}{P} \right)  \; . 
\end{equation}
In our case, though, $\Gamma_{\rm FKV}=\Gamma(X_{\rm FKV})$ turns out to be very involved analytically and must be evaluated numerically.

\begin{figure*}[t]
\begin{center}
\hbox{\includegraphics[width=0.5\textwidth]{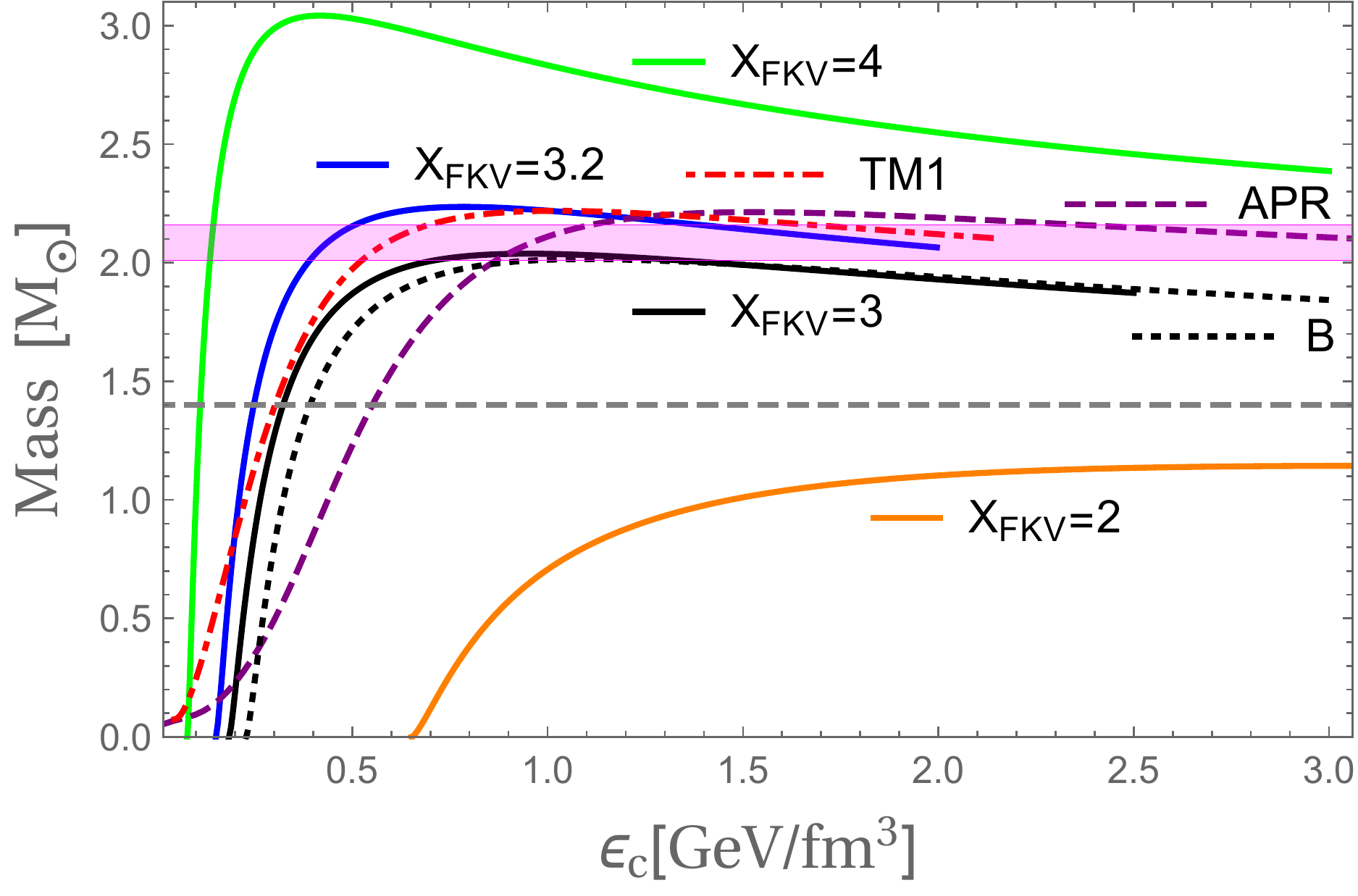}
	  \includegraphics[width=0.5\textwidth]{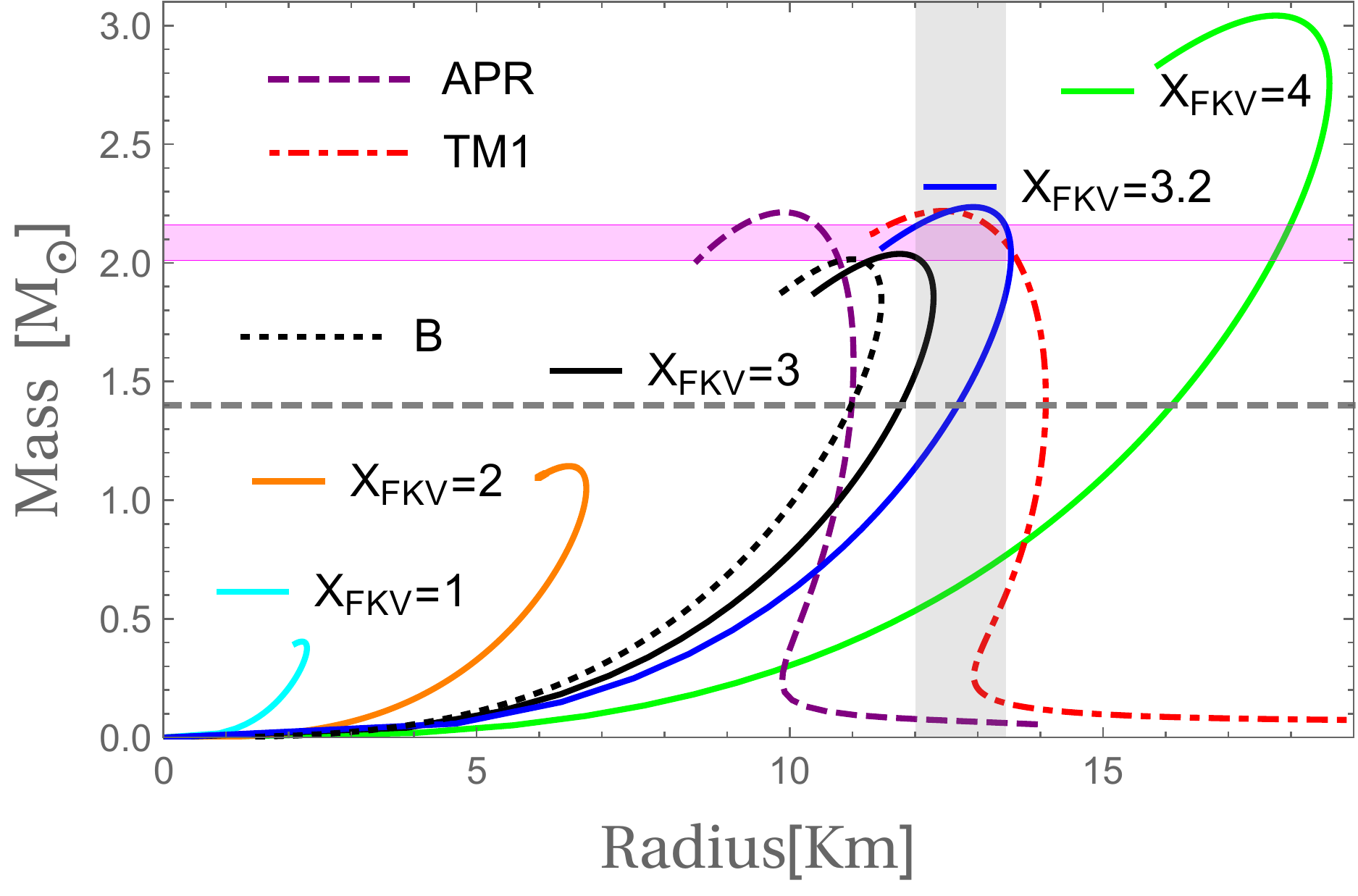}}
\caption{\textit{Left panel:} Total gravitational mass $M$ (in solar mass units) versus central energy density, $\epsilon_{c}$, for the same EoSs of Fig. \ref{fig:AnomAndEoSs}. The horizontal light-purple band represents the astrophysical constraint coming the gravitational wave signal GW170817 \cite{Rezzolla:2017aly}. The horizontal gray dashed line represents stellar configurations with $M=1.4M_\odot$. 
\textit{Right panel:} Mass-radius diagram for EoSs from pQCD for a few values of $X_{\rm FKV}$, the bag model $B$, and nuclear matter EoSs APR and TM1 (see text). The vertical light-gray band represents the maximum (13.45km) and minimal (12km) radii for a NS with $M=1.4M_\odot$ when using the gravitational wave constraint of Ref. \cite{Most:2018hfd}.}
\label{fig:MassDensRad}
\end{center}
\end{figure*}

\section{Results and discussion}
	\label{sec:QManalysis}		

We can now investigate the behavior of different eigenfrequencies, $\omega_{n}$, of the fundamental and first excited modes produced by a radial perturbation in a quark (or strange) star making use of the framework built in Sec. \ref{sec:setup}.	

\subsection{Hydrostatically equilibrated quark stars}
	
We first solve the TOV equations for the perturbative QCD EoS using the FKV formula for some values of the renormalization scale $X_{\rm FKV}$. For the sake of comparison, we also display results obtained for the nuclear matter EoSs mentioned in Sec. \ref{sec:setup}. 

In Fig. \ref{fig:MassDensRad} we show our results for the mass as a function of the central energy densities and in the mass-radius diagram. In both panels we indicate the astrophysical constraint on the maximum mass of NS obtained from the gravitational waves coming from the merger event GW170817, namely between $2.01$ and $2.16$ solar masses \cite{Rezzolla:2017aly} (horizontal light-purple band). This event additionally puts a constraint on the radius of a NS of $1.4 M_\odot$ to be between 12 and 13.45 km \cite{Most:2018hfd}, which we indicate in the right panel of this figure as a vertical gray band. From this panel it is straightforward to see that only values of $X_{\rm FKV}$ between 3 and 3.2 satisfy simultaneously the GW170817 constraints of mass and radius, whereas the APR and TM1 EoSs nearly satisfy the mass constraint but not the radius restriction.

To make our discussion more quantitative, we show a few illustrative tables. In Table \ref{tab:table1} we list the associated values of minimal (at the quark star surface) and maximal (at quark star center) baryon chemical potentials, $\mu^{\rm (min,max)}_{B}$, corresponding to the star with maximum mass. We also present values of the associated central energy densities $\epsilon^{\rm max}_{c}$ and radii $R^{\rm max}$. Notice that the maximal values of $\mu_{B}$ for the APR and TM1 EoSs lie slightly above the quark analogues. However, the APR EoS violates the causal limit before reaching its maximum mass configuration listed in Table \ref{tab:table1}. Moreover, it was estimated in Ref. \cite{Lattimer:2010uk} that the maximal value of baryon chemical potential at the center of NS would be $2.1$ GeV. 
\begin{table*}[t]
  \begin{center}
    \begin{tabular}{c|c|c|c|c|c} 
       ${\rm EoS}$ & $\mu^{\rm min}_{B}[{\rm GeV}]$ & $\mu^{\rm max}_{B}[{\rm GeV}]$ & $\epsilon^{\rm max}_{c} [{\rm GeV/fm}^3]$ & $M^{\rm max}[{\rm M_{\odot}}]$ & $R^{\rm max}[{\rm km}]$\\
      \hline
        $1$ 		&$2.01$ &$3.091$ &$26.29$ & $0.404087$ & $2.212$\\
        $2$ 		& $1.21$&$1.8376$ &$3.1$ & $1.14363$ & $6.47548$\\
        $3$ 		& $0.91248$ &$1.39$ &$0.982$ & $2.03809$ & $11.7532$\\
        $3.2$ 	& $0.87251$ &$1.323$ &$0.8$ & $2.23551$ & $12.9284$\\
        $4$ 		& $0.75267$ &$1.13$ &$0.416$ & $3.04224$ & $17.757$\\
        $B$ &$0.8285$&$1.2981$ &$1.0977$ & $2.02$ & $10.99$\\
        APR     & $0.9268$ &$2.269$ &$1.5337$ & $2.2$ & $10$\\
        TM1     & $0.932276$ &$1.628$ &$1.02$ & $2.2$ & $13.5$\\
        \end{tabular}
        \caption{Equations of state from pQCD (for which we only show the value of $X_{\rm FKV}$), the bag model, and nuclear matter (APR and TM1); minimal and maximal baryon chemical potentials; central energy densities for the maximum mass configurations; maximal masses; and corresponding radii.}
      \label{tab:table1}
  \end{center}
\end{table*}
%

\begin{figure*} [t]
\begin{center}
\hbox{\includegraphics[width=0.5\textwidth]{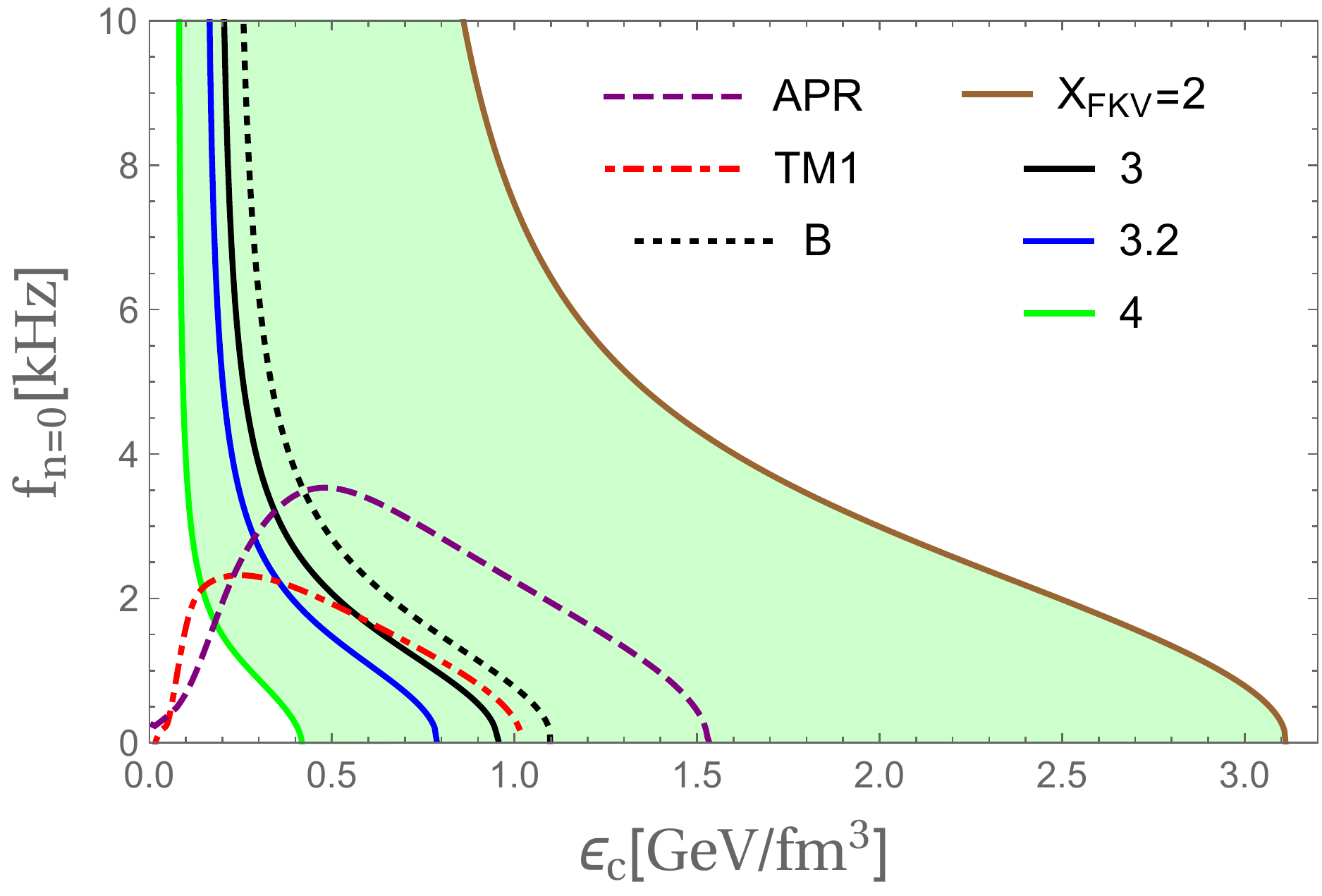}
	  \includegraphics[width=0.5\textwidth]{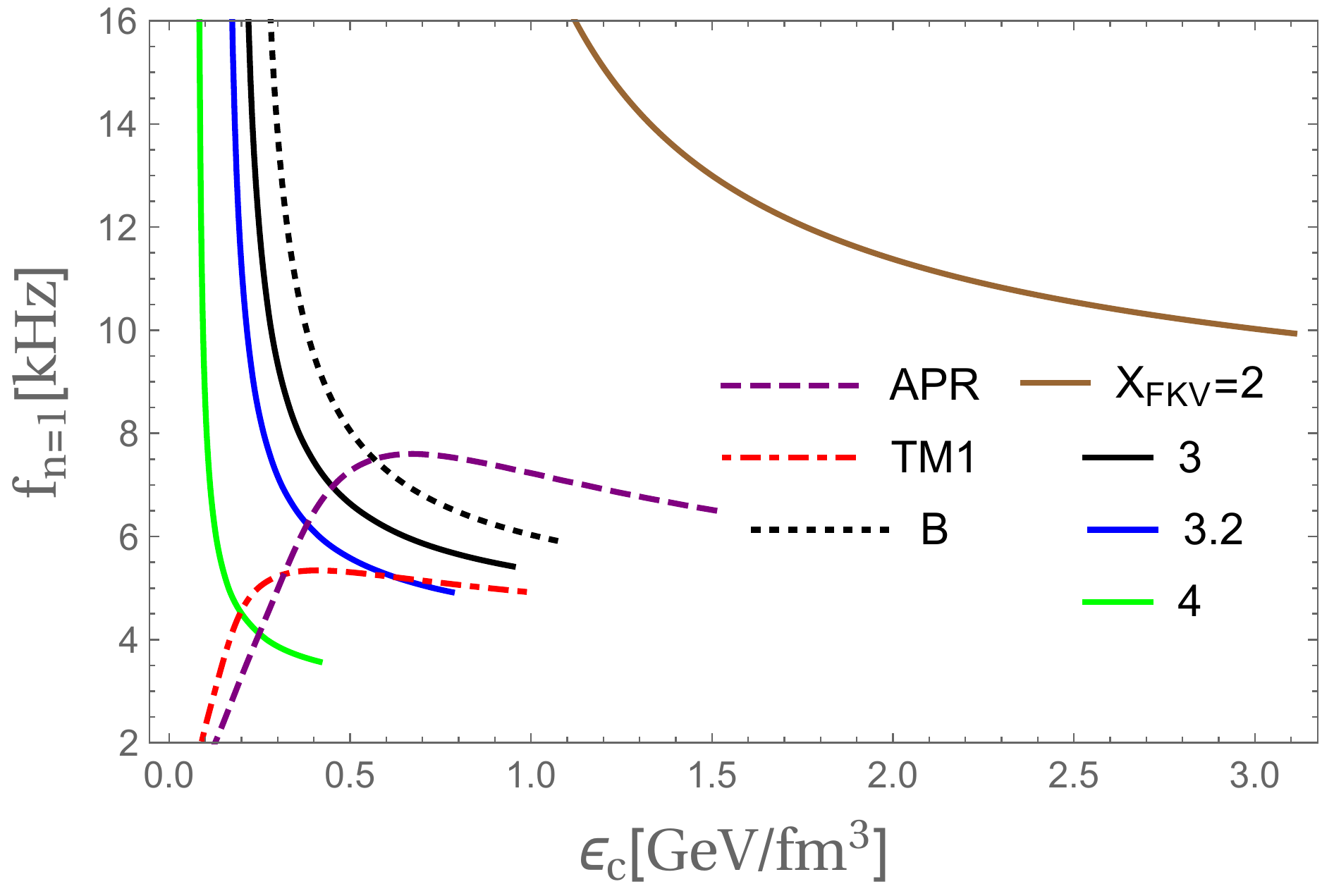}}
\hbox{\includegraphics[width=0.5\textwidth]{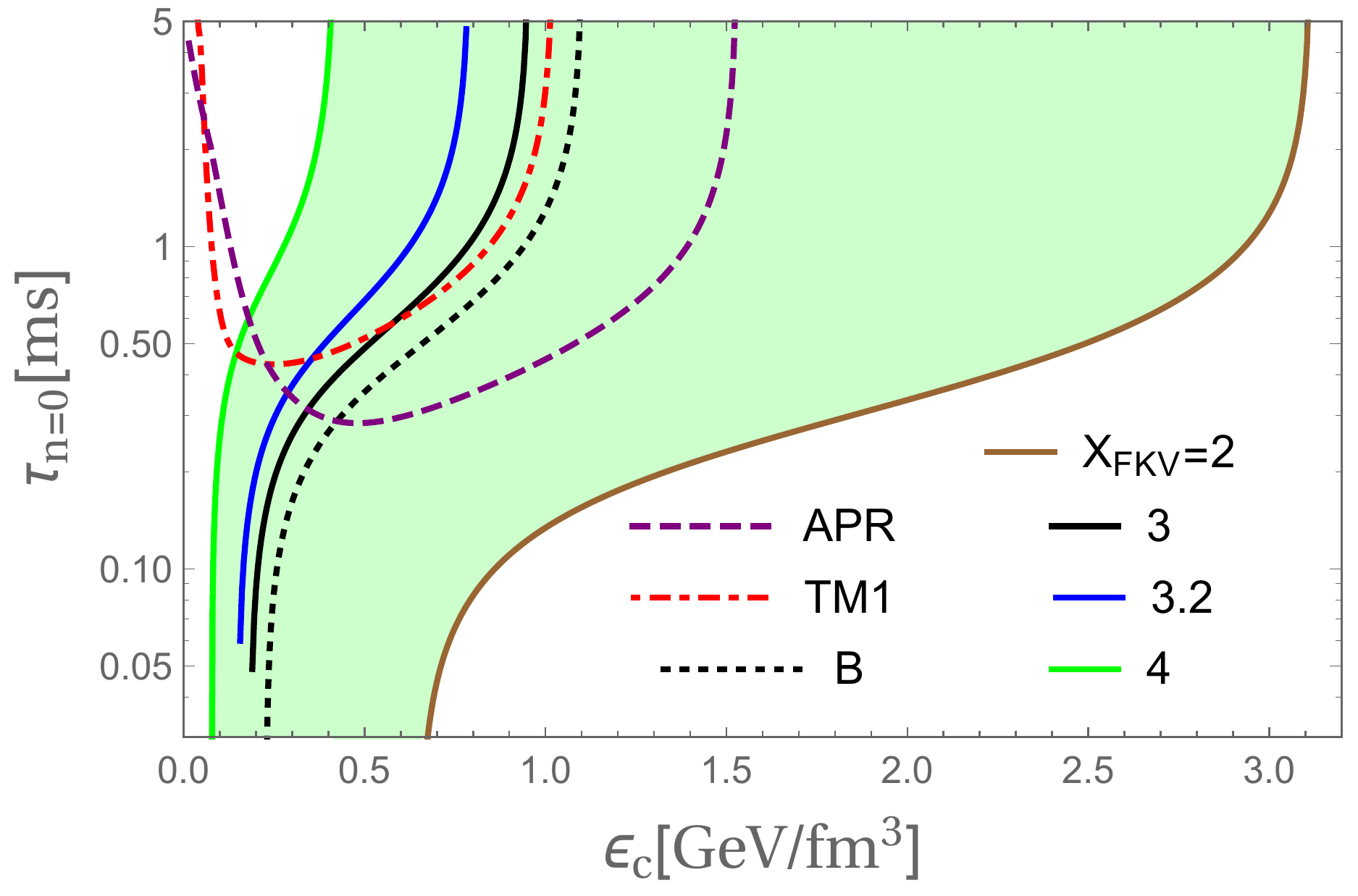}
	  \includegraphics[width=0.5\textwidth]{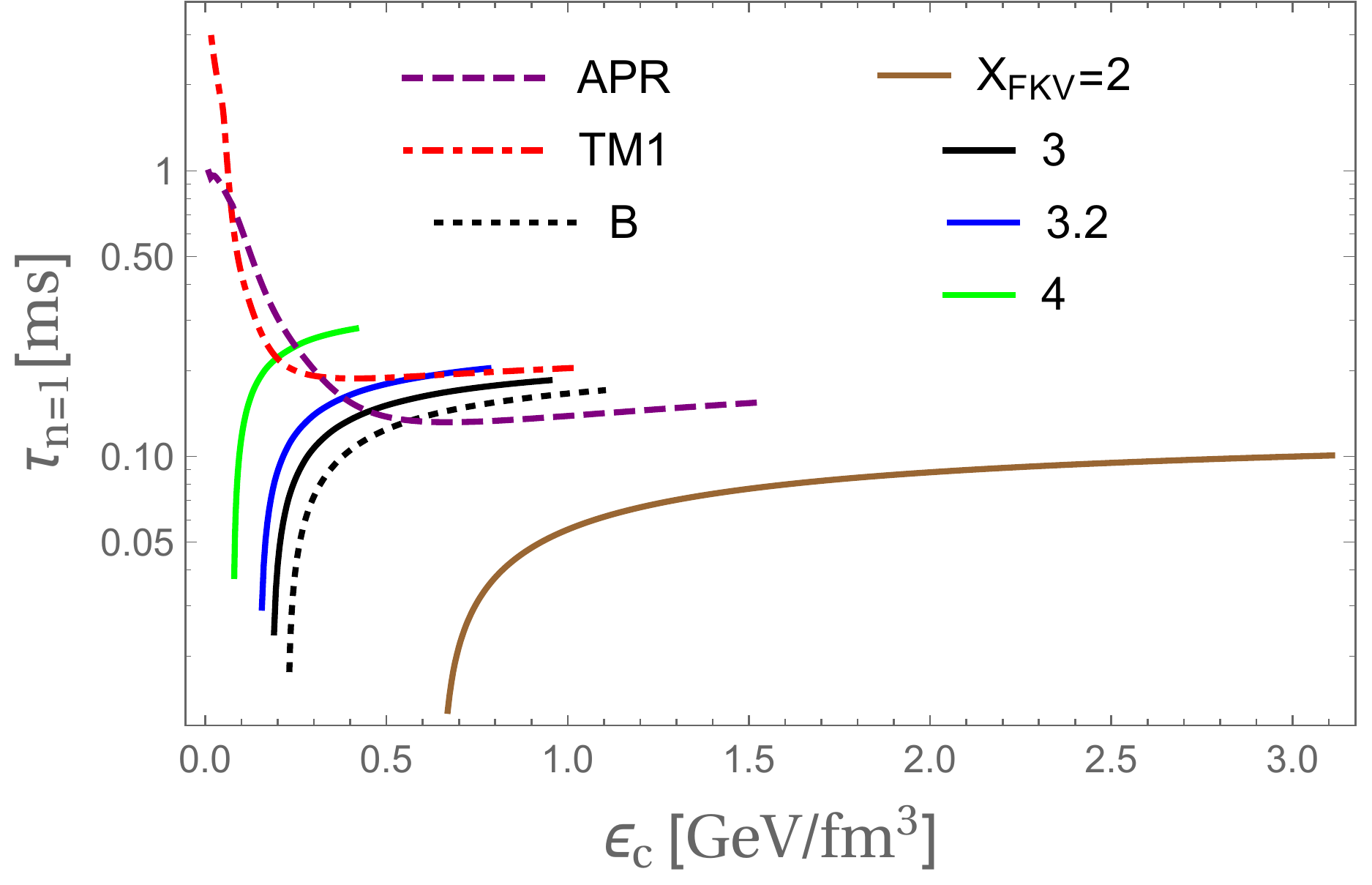}}
\caption{\textit{Upper panels}: Fundamental mode, $f_{n=0}$, and first mode, $f_{n=1}$, frequencies as functions of the central energy density. \textit{Lower panels:} Same for the periods. We show results using EoSs from pQCD, the bag model ($B$) and nuclear matter (APR and TM1).}
\label{fig:RadOscDensity}
\end{center}
\end{figure*}

\subsection{Dynamically stable quark stars}
	
The necessary condition for stability of compact stars requires that the stellar configurations, obtained after solving the TOV equations, satisfy the condition $\partial{M}/\partial\epsilon_{c}\geq{0}$. The behavior for the quark stars obtained from pQCD is shown in the left-panel of Fig. \ref{fig:MassDensRad}. However, one still has to solve the pulsation equations to guarantee their stability, so we must study their associated eigenfrequencies. 

For simplicity, in the following we write the eigenfrequencies, $\omega_{n}$, in terms of the linear frequency defined as $f_{n}~{\equiv}~{\omega_{n}/{2\pi}}$. In particular, the fundamental and first excited oscillation modes, i.e., $n=0,1$, are very relevant since they are the easiest to be excited by external (radial) perturbations. Besides, they will turn out to be very sensitive to the interactions encoded in the EoS from high-density perturbative QCD. Higher eigenfrequencies ($n=2,3,...$) can also be calculated being apparently interesting for potential observations since they have larger numerical values. However, they are too difficult to be excited in realistic situations and thus we do not exhibit their values in this work. 

\begin{figure*} [t]
\begin{center}
\hbox{\includegraphics[width=0.5\textwidth]{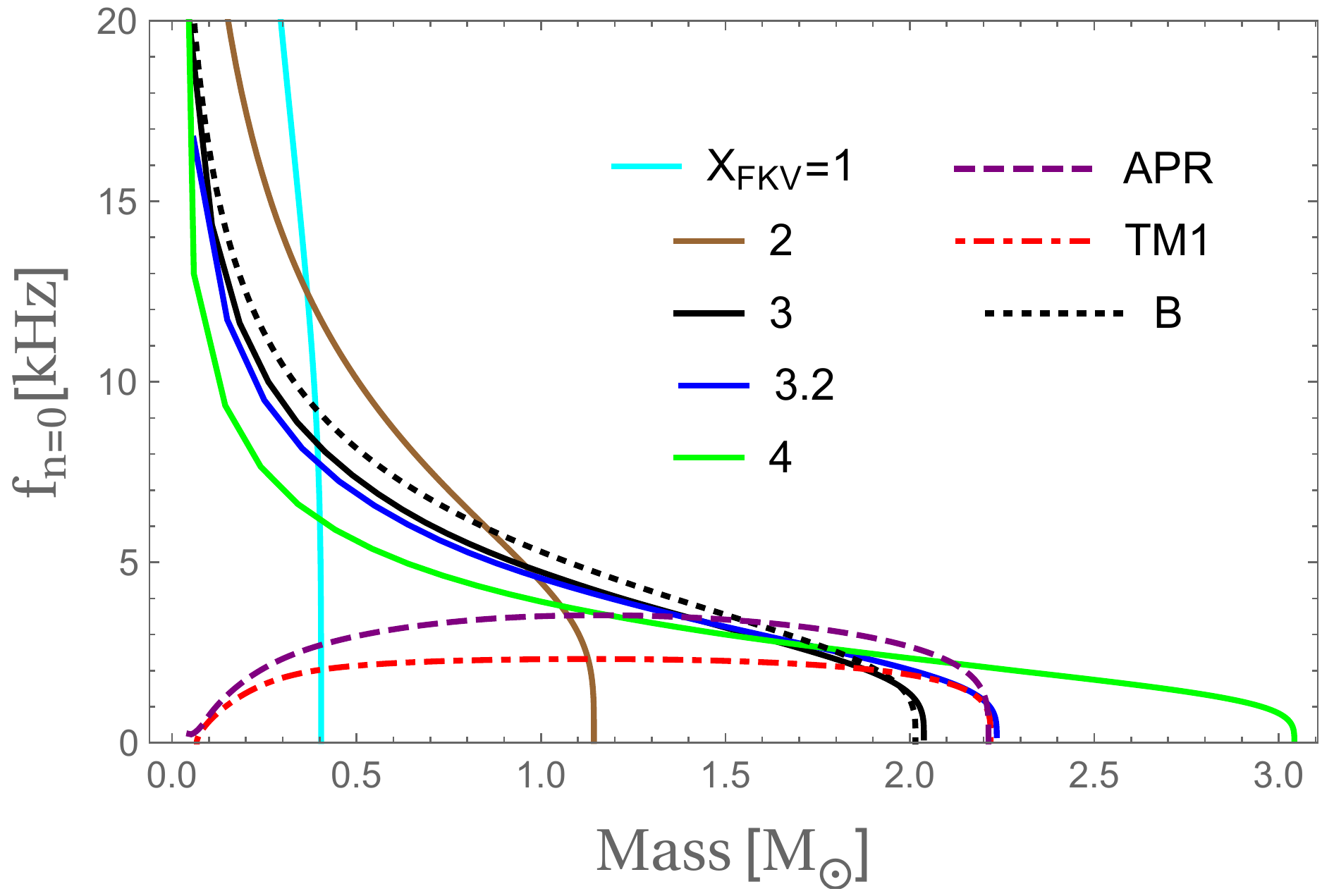}
	  \includegraphics[width=0.5\textwidth]{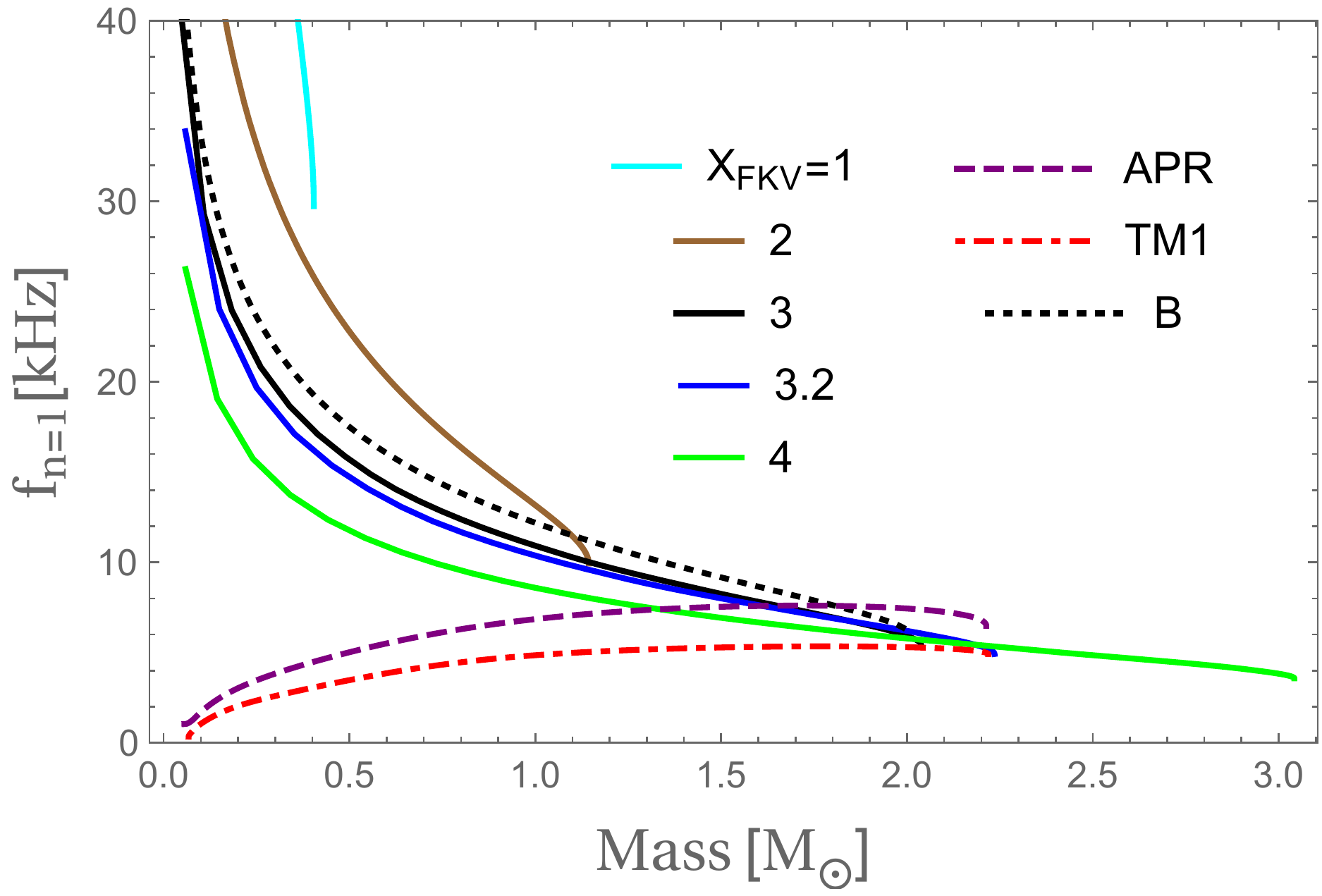}}
\hbox{\includegraphics[width=0.5\textwidth]{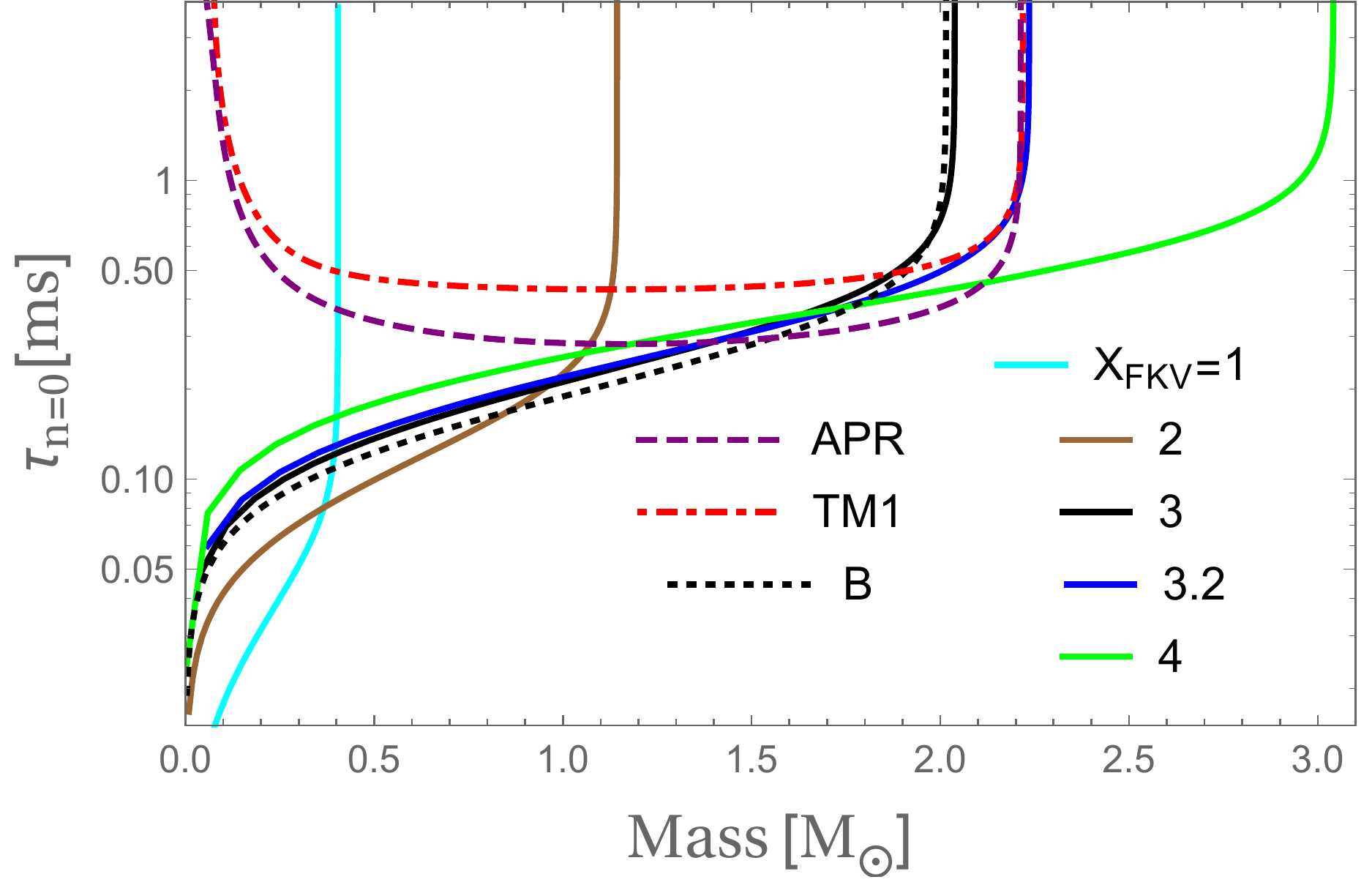}
	  \includegraphics[width=0.5\textwidth]{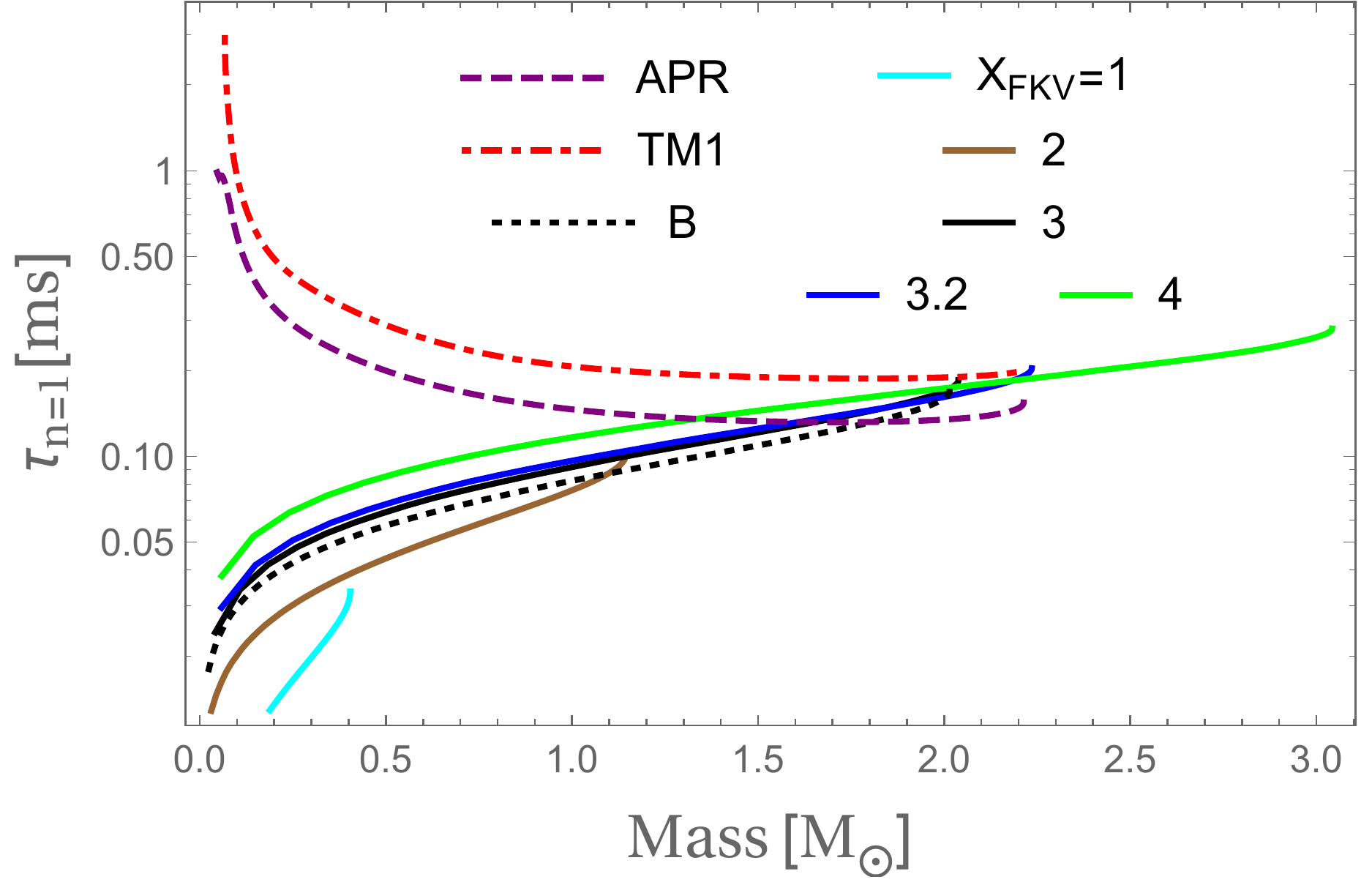}}
\caption{Same as in Fig. \ref{fig:RadOscZ} but now the frequencies and periods are functions of the total gravitational mass $M$.}
\label{fig:RadOscMass}
\end{center}
\end{figure*}

\begin{figure*} [t]
\begin{center}
\hbox{\includegraphics[width=0.5\textwidth]{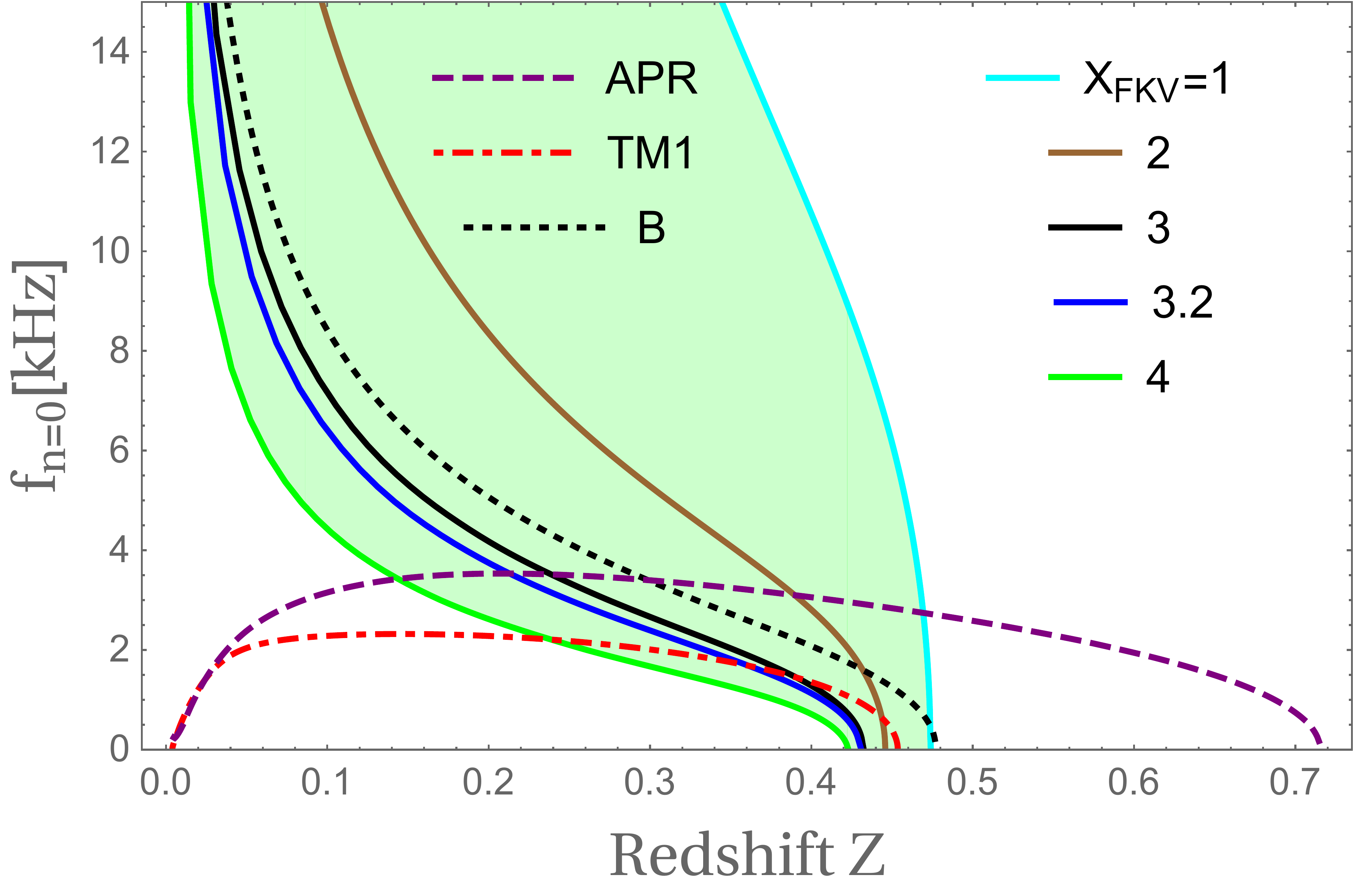}
	  \includegraphics[width=0.5\textwidth]{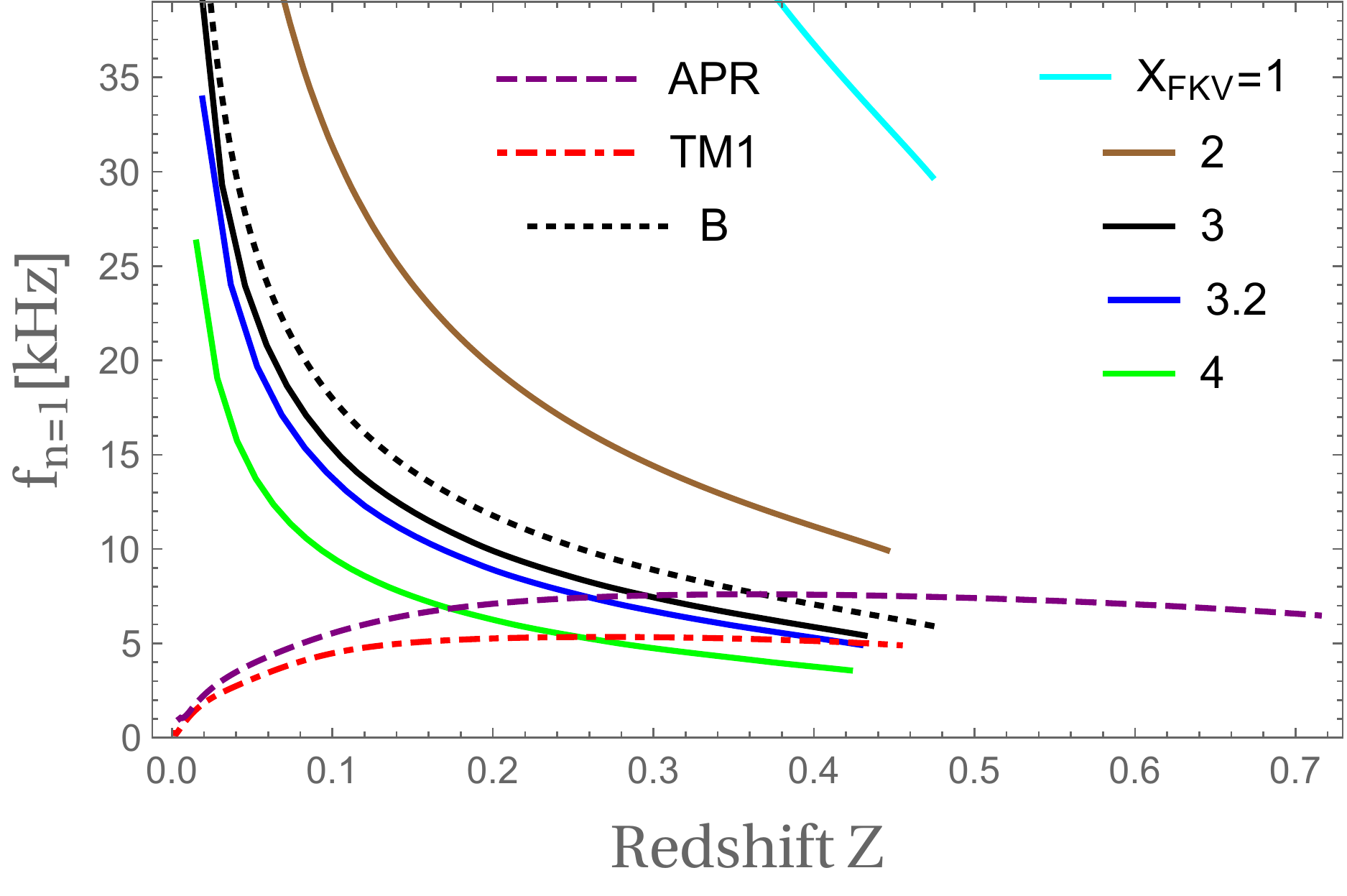}}
\hbox{\includegraphics[width=0.5\textwidth]{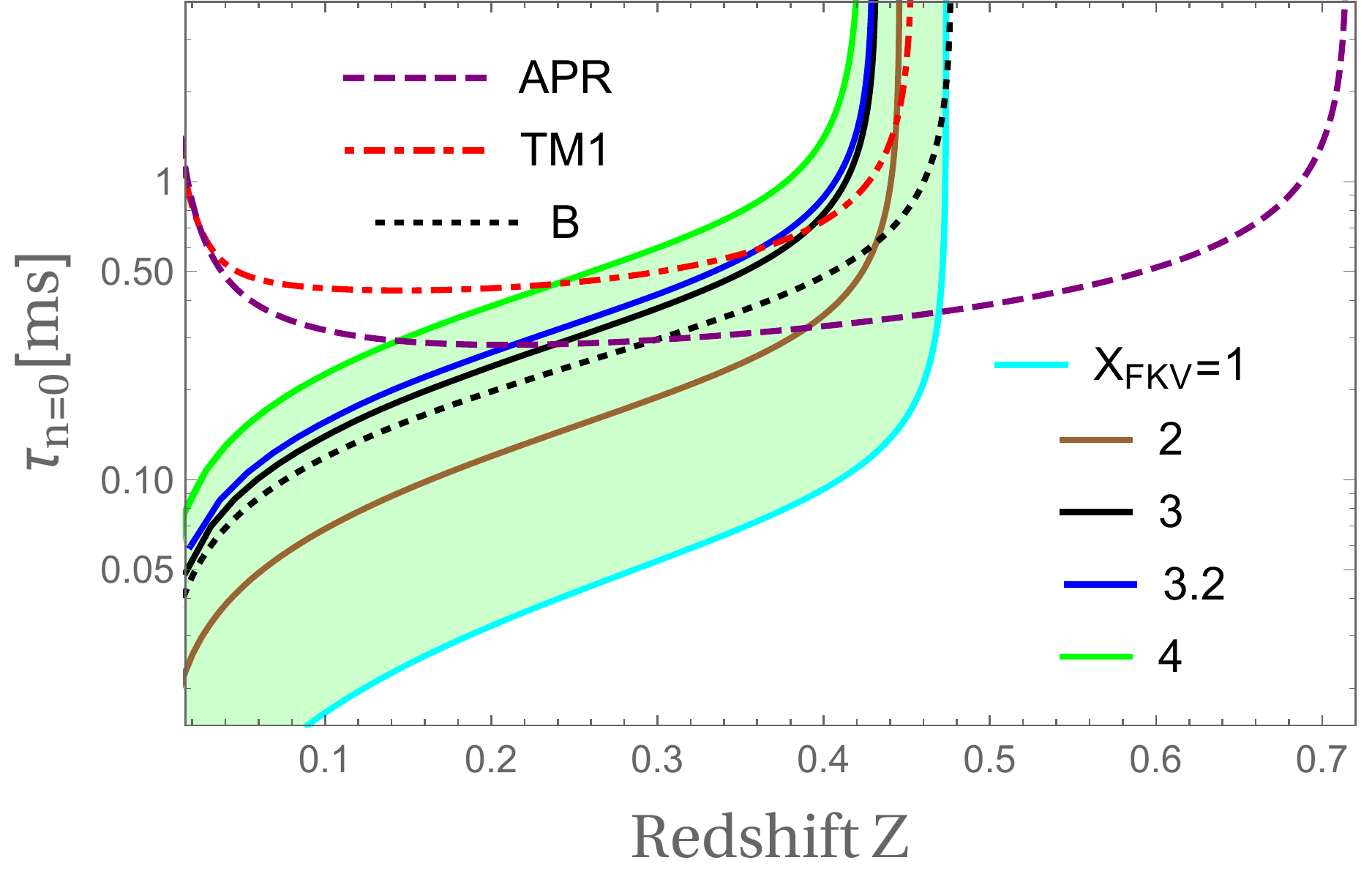}
	  \includegraphics[width=0.5\textwidth]{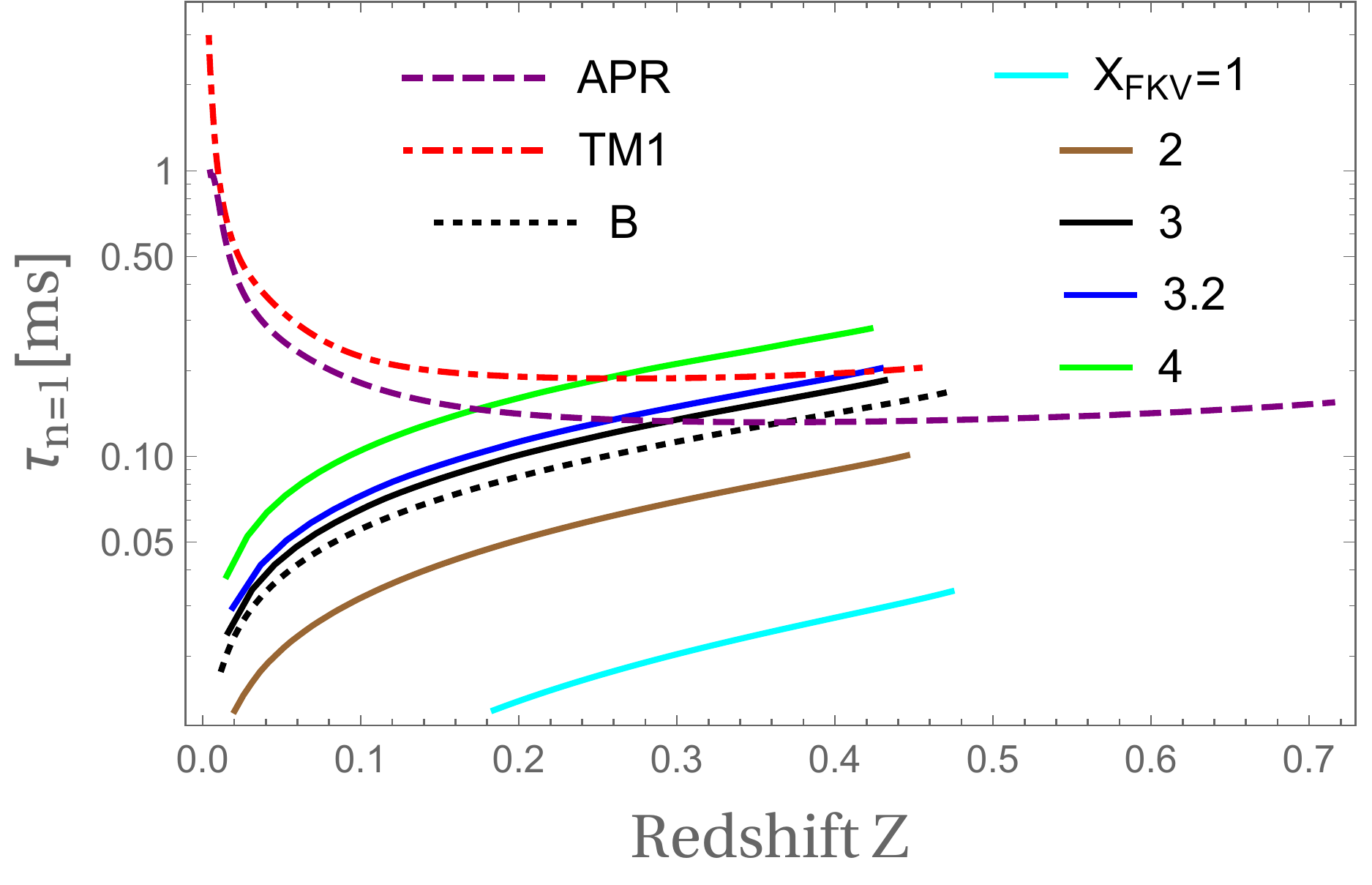}}
\caption{Same as in Fig. \ref{fig:RadOscZ} but now the frequencies and periods are functions of the redshift parameter Z.}
\label{fig:RadOscZ}
\end{center}
\end{figure*}

From Figs. \ref{fig:RadOscDensity}, \ref{fig:RadOscMass} and \ref{fig:RadOscZ}, we can see that the fundamental ($n=0$) and first-excited ($n=1$) mode frequencies, $f_{n=0}$ and $f_{n=1}$ respectively, behave differently for different values of $X_{\rm FKV}$, producing a large band of possibilities. As expected, their behavior is quite different from the nuclear matter EoSs (APR and TM1) which we plot only for comparison. It is clear from these figures that different renormalization scales affect qualitative features for these radial oscillation frequencies. Moreover, the scaling law for the periods of the bag model (and some of its modified versions), $\bar{\tau}_{n}=(B/\bar{B})^{1/2}\times{\tau}_{n}$ 
\cite{Benvenuto:1991}, is not realized in the case of the equation of state coming from cold and dense perturbative QCD since conformal invariance is broken by interactions via the running of the strong coupling and quark masses. In Fig. \ref{fig:RadOscZ}) we show the dependence of frequencies and periods on redshift parameter $Z=(1-2M/R)^{-1/2}-1$ \cite{Glendenning:2000}. This can be useful since allows us to compare two observable quantities in astronomical measurements.

Let us now turn our attention to the dependence of the frequencies and periods on the central energy density, gravitational mass and redshift.

\subsubsection{Dependence on the energy density}

Fig. \ref{fig:RadOscDensity}, shows that, although the quantitative behavior of the fundamental and first excited ($n=0,1$) vibrational modes (as functions of $\epsilon_{c}$) of for quark stars are very sensitive to $X_{\rm FKV}$, their qualitative behavior is similar to the bag model (${\rm B^{1/4}=145MeV}$) for all values of central energy density. As expected, nuclear matter stars (APR and TM1) behave quite differently. In Table \ref{tab:table2} we list values of central energy densities, frequencies and periods for the modes $n=0,1$ for a canonical neutron star of $M=1.4M_{\odot}$ for the EoSs we use along this work. The choice of mass follows from the fact that most of the observed pulsars tend to have masses near this value. Notice that although the bag model surpasses the two-solar mass constraint only for high central energy densities, the EoS obtained from the FKV formula requires relatively low-energy central densities to produce heavy stars. 

%
\begin{table*}[t]
  \begin{center}
    \begin{tabular}{c|c|c|c|c|c} 
      ${\rm EoS}$ & $\epsilon_{c}[{\rm GeV/fm^{3}}]$ & $f_{0}[{\rm kHz}]$ & $\tau_{0}[{\rm ms}]$ & $f_{1}[{\rm kHz}]$ & $\tau_{1}[{\rm ms}]$\\
      \hline
      $3$ & $ \approx 0.30$ &$ \approx 3.5 $ &$ \approx 0.29 $ & $\approx 9$ & $\approx 0.11$ \\
      $3.2$ & $ \approx 0.26$ &$ \approx 3.5 $ &$ \approx 0.29 $ & $\approx 9$ & $\approx 0.11$ \\
      $4$ & $ \approx 0.10$ &$ \approx 3.3 $ &$ \approx 0.30 $ & $\approx 8$ & $\approx 0.13$ \\
      ${\rm B}$ & $ \approx 0.40$ & $ \approx 4.0 $ &$ \approx 0.25 $ & $\approx 10$ & $\approx 0.10$ \\
      APR & $ \approx 0.56$ & $ \approx 3.5$ &$ \approx 0.29$ & $\approx 8$ & $\approx 0.13$ \\
      TM1 & $ \approx 0.29$ & $ \approx 2.0 $ &$ \approx 0.50 $ & $\approx 6$ & $\approx 0.17$ \\
    \end{tabular}
        \caption{
Values of central energy densities ($\epsilon_{c}$), fundamental ($f_{n=0}$) and first-excited ($f_{n=1}$) mode frequencies, and their associated periods ($\tau_{n=0}$ and $\tau_{n=1}$, respectively) for stars with mass $M=1.4M_{\odot}$, obtained from equations of state from the FKV formula (for different values of $X_{\rm FKV}$), for nuclear matter and the bag model (see text).}
    \label{tab:table2}
  \end{center}
\end{table*}
\begin{table*}[t]
  \begin{center}
    \begin{tabular}{c|c|c|c|c} 
       ${\rm EoS}$ & $f^{\rm lower}_{0}[{\rm kHz}]$ & $\tau^{\rm lower}_{0}[{\rm ms}]$ & $f^{\rm upper}_{0}[{\rm kHz}]$ & $\tau^{\rm upper}_{0}[{\rm ms}]$ \\
      \hline
        $3$ & $ \approx 1.0$ &$ \approx 1.0 $ & - & -\\
        $3.2$ & $ \approx 2.0$ &$ \approx 0.5 $ &$ \approx 1.0 $ & $\approx 1.0$\\
        $B$ & $ \approx 1.0$ &$ \approx 6.0^{(*)} $ & - & -\\
        APR & $ \approx 3$ &$ \approx 0.33 $ &$ \approx 2.0 $ & $\approx 0.5$\\
        TM1 & $ \approx 2.0$ &$ \approx 0.5 $ &$ \approx 1.0 $ & $\approx 1.0$\\
        \end{tabular}
        \caption{Frequencies and periods of the fundamental mode for stellar configurations satisfying the gravitational wave event GW170817 on compact stars as having masses between $M^{\rm lower}_{\rm max}=2.01M_{\odot}$ and $M^{\rm upper}_{\rm max}=2.16M_{\odot}$. The period marked with (*) is notably different from the ones for other stellar configurations because it is very close to the maximum mass where the fundamental period diverges.}
      \label{tab:table3}
  \end{center}
\end{table*}
\begin{table*}[t]
  \begin{center}
    \begin{tabular}{c|c|c|c|c} 
       ${\rm EoS}$ & $f^{\rm lower}_{1}[{\rm kHz}]$ & $\tau^{\rm lower}_{1}[{\rm ms}]$ & $f^{\rm upper}_{1}[{\rm kHz}]$ & $\tau^{\rm upper}_{1}[{\rm ms}]$ \\
      \hline
        $3$ & $ \approx 6.0$ &$ \approx 0.17 $ & - & -\\
        $3.2$ & $ \approx 6.0$ &$ \approx 0.17 $ &$ \approx 5.0 $ & $\approx 0.20$\\
        $B$ & $ \approx 6.0$ &$ \approx 0.17 $ & - & -\\
        APR & $ \approx 8.0$ &$ \approx 0.13 $ &$ \approx 8.0 $ & $\approx 0.13$\\
        TM1 & $ \approx 5.0$ &$ \approx 0.20 $ &$ \approx 5.0 $ & $\approx 0.20$\\
        \end{tabular}
        \caption{Same as Table \ref{tab:table3} but for frequencies and periods of the first-excited mode of oscillation for the mentioned EoSs.}
      \label{tab:table4}
  \end{center}
\end{table*}

\subsubsection{Dependence on the gravitational mass}

It is clear from Figs. \ref{fig:MassDensRad} and \ref{fig:RadOscMass} that choosing $X_{\rm FKV}\gtrsim{1}$ yields very compact quark stars with higher (lower) values of frequencies (periods), in contrast to the ones provided by the bag model (the opposite happening for larger values of $X_{\rm FKV}$). For instance, the fundamental period of $X_{\rm FKV}=1$ takes a maximum value of approximately 0.1 milliseconds before it diverges at its maximum mass configuration. Notice also that for $X_{\rm FKV}$ approximately between $3$ and $4$, although producing heavy strange quark stars, satisfying the two-solar mass constraint straightforwardly, their sector of low-mass strange quark stars have lower values of frequency signalling that strong interactions may play a role in making those stars less deformable, i.e. less compact, against external radial perturbations\footnote{The frequencies and periods were calculated for $X_{\rm FKV}\sim3(3.2)$ that generate maximum masses of $\sim2(2.2)M_{\odot}$, respectively.}.  

Taking into account the recent gravitational wave constraint from the GW170817 event on the maximum gravitational mass of neutron stars as being in the range $2.01M_{\odot} \leq M_{\rm max} \lesssim 2.16M_{\odot}$ \cite{Rezzolla:2017aly,Most:2018hfd}, we can extract additional limits on the values of oscillation frequencies and periods. We list the values of fundamental (first-excited) oscillation frequencies and corresponding periods in Table \ref{tab:table3} (\ref{tab:table4}) for stellar configurations within this range of maximum mass, indicating the values corresponding to the lower and upper limits in the previous inequality.

Strange stars with masses around the $2M_{\odot}$ limit have periods that tend to be higher than $1$ ms, whereas low-mass strange stars tend to have periods that are smaller and smaller, making them difficult to detect by modern techniques including drifting subpulses and micropulses \cite{Benvenuto:1991jz}. The value of $X_{\rm FKV}$ is then constrained to be in the range of $\sim 3-3.2$. The period for the case of quark stars tend to be in the range of $\sim 0.4-2.9$ ms, which is something new from pQCD that the bag model $B$ cannot reproduce since although it can reach two solar masses, it cannot go above this limit without violating the Bodmer-Witten hypothesis unless effective interaction terms are added to the equation of state \cite{Weissenborn:2011qu}.

\subsubsection{Dependence on the redshift}

The dependence of the frequencies and periods of quark stars on the gravitational redshift parameter Z are displayed in Fig. \ref{fig:RadOscZ}. From this figure, it becomes clear that, independently of the particular EoS used (for different values of $X_{\rm FKV}$ in pQCD, $B$ and TM1) and their maximum masses, the maximum gravitational redshift $Z$ tends to accumulate in the region between 0.42 and 0.48, which can be used to restrict the behavior of the EoS for dense matter when compared to current astronomical observations of $Z$. Although the APR case lies outside this region, one should recall that at high densities (before reaching its maximal mass configuration) it becomes superluminal. Notice that the first-excited mode displayed in Fig. \ref{fig:RadOscZ} seems to distinguish low-mass quark stars from purely hadronic stars.


\section{Summary and final remarks}
  \label{sec:conclusion}  
  
In this paper we have investigated the relativistic radial oscillations of unpaired bare quark stars and strange stars using an equation of state from perturbative QCD, including up, down, and strange quarks in a cold, dense medium in $\beta$-equilibrium and electrically neutral. For the best of our knowledge, similar studies of the radial oscillation stability were only performed within the MIT bag model framework (occasionally including  minor modifications). Our results contains a natural estimate of the inherent systematic uncertainties in the evaluation of the equation of state, and therefore of all observables that follow, and might bring new insights into the phenomenology of quark stars and their possible observational searches.

Comparing the nucleonic and quark star results obtained in this work, one finds that their fundamental and first excited modes are quite distinguishable for low-mass stars. On the other hand, heavy stars become numerically indistinguishable in the region near the two-solar mass limit. Nevertheless, their curves are different and this could be important to map their vibrational behavior. In fact, this could be used to discriminate between hadronic and quark stars by comparing their non-radial pulsation modes (which are correlated to the radial modes in gravitational waves \cite{Passamonti:2005cz,Passamonti:2007tm}), especially in the case of heavy compact stars, close to the current constraint on their maximum mass \cite{Flores:2013yqa,Flores:2018pnn}.  
 
Our results represent an initial step towards the more realistic case of hybrid star pulsations (see, e.g., Ref. \cite{Sahu:2001iv}), where a hadronic mantle and crust effects should be included \cite{future}. They might also shed light onto the phenomenology of strange dwarfs, which seem to be unstable under radial  perturbations \cite{Glendenning:1994zb,Alford:2017vca}, or more exotic forms including the existence of condensed dark matter in neutron stars \cite{Deliyergiyev:2019vti} and strange stars \cite{Panotopoulos:2017eig}.


\begin{acknowledgments}
We thank J\"{u}rgen Schaffner-Bielich for useful comments. This work was partially supported by CAPES (Finance Code 001), CNPq, FAPERJ, and INCT-FNA (Process No. 464898/2014-5).
\end{acknowledgments}



\end{document}